\documentclass{article}

\usepackage{bm,amsmath,amsthm,amssymb}

\theoremstyle{plain}

\theoremstyle{definition}

\theoremstyle{remark}

\numberwithin{equation}{section}

\def\bu{{\bm u}}

\def\bz{{\bm z}}

\def\bmf{{\bm f}}

\begin{document}

\title{A Note on Machine Learning Approach for Computational Imaging}
\author{Bin Dong \\Beijng International Center for Mathematical Research,\\ Peking University}
\date{}


\maketitle

\begin{abstract}
Computational imaging has been playing a vital role in the development of natural sciences. Advances in sensory, information, and computer technologies have further extended the scope of influence of imaging, making digital images an essential component of our daily lives. For the past three decades, we have witnessed phenomenal developments of mathematical and machine learning methods in computational imaging. In this note, we will review some of the recent developments of the machine learning approach for computational imaging and discuss its differences and relations to the mathematical approach. We will demonstrate how we may combine the wisdom from both approaches, discuss the merits and potentials of such a combination and present some of the new computational and theoretical challenges it brings about.
\end{abstract}

\section{Introduction}\label{S:ML}

The development of natural sciences has been heavily relying on visual examinations. Therefore, images play a central role since they can accurately record the phenomenon of interest and be further processed and analyzed by algorithms to assist human decision-making. Advances in sensory, information, and computer technologies have made it possible to apply some of the most sophisticated developments in mathematics and machine learning to the design and implementation of efficient algorithms to process and analyze image data. As a result, the impact of images has now gone far beyond natural sciences. Image processing and analysis techniques are now widely adopted in engineering, medicine, technical disciplines, and social media, and digital images have become an essential element of our daily lives.

The term ``computational imaging'' often refers to the process of forming images from measurements via computations. It consists of two crucial steps, sensing (or scanning) with a detector or electromagnetic beam and image reconstruction with a numerical algorithm. In this note, we shall extend the definition of computational imaging to include image analysis (or interpretation). Sensing is the most fundamental step of computational imaging, where hardware design is the central research topic assisted by sampling theory and algorithms (e.g., compressed sensing \cite{candes2006robust,donoho2006compressed}). Its main objective is to acquire high-quality measurements of the imaging subject efficiently. Image reconstruction is a mid-level task in computational imaging that comes after the step of sensing. Its main objective is to reconstruct high-quality images from measurements. Mathematics has been the main driven force in the advancement of image reconstruction for the past few decades \cite{aubert2006mathematical,ChanShen,Dong2010IASNotes}.
Conversely, image reconstruction also brings to mathematics new challenging problems and fascinating applications that gave birth to many new mathematical tools, whose application has even gone beyond the scope of image reconstruction. Image analysis is a high-level task in computational imaging where both mathematical and statistical models have played a significant role \cite{mumford2010pattern,soille2013morphological}. The main objective of image analysis is to extract meaningful information from images to assist in human decision-making. 

Although mathematical and statistical models have been successful in computational imaging, they also face many challenges. In the sensing step, although the theory of compressed sensing tells us random sampling is a good option for some instances, only a handful of imaging modalities (e.g., MRI) satisfy the theory's assumptions. Furthermore, a better sensing mechanism should be adaptive to each subject that the theory of compressed sensing is not considered. A good adaptive sensing mechanism needs to decide for each given imaging subject on which measurements to take to maximize a particular quality metric. For image reconstruction, existing models and algorithms are designed based on human knowledge. Although we know for each model which general class of images is most suitable, e.g., we know TV model \cite{ROF} is ideal for piecewise constant images, for a given set of natural images, it is hard to design a most suitable model entirely by hand (e.g., the design of the regularization, the hyperparameters, etc.). For image analysis, the ultimate objective is to extract relevant image features to facilitate decision-making. Before deep learning, features are often handcrafted, which may not be well adaptive to the data set or the underlying image analysis task. For example, we know image edges, textures and contents are essential for image segmentation. However, it is unclear what contents mean for different images and whether there are other important features for image segmentation.

These challenges that limit the further development of computational imaging can be rephrased in a mathematical term as the challenge of approximating high-dimensional (HD) nonlinear functions of which we have limited or no knowledge, primarily due to the well-known curse of dimensionality. For example, making decisions on the set of measurements optimal for a given imaging subject can be an intricate HD function that takes the current state (e.g., the existing measurements) as the input and the next set of measurements as the output; deciding on the best hyperparameters (e.g. the sparsifying transformation, the sparsity promoting norm, the regularization parameters, etc.) of image reconstruction models for a given data set is another example; the embedding of images to the feature space (or latent space) to facilitate different image analysis tasks can also be a complicated HD function. This challenge is now being overcome by deep learning to various extents \cite{deep-learning-book-2016}. Deep neural networks (DNNs), especially the convolutional neural networks (CNNs), are very effective in approximating nonlinear HD functions. Therefore, DNNs are now widely used in computational imaging, and the field has advanced significantly over the past decade. We shall review some of these exciting advancements, suggest some future opportunities, and raise some new theoretical challenges.

\section{Improving Image Reconstruction with Deep Learning}\label{SS:Unrolling}

A typical learning-based image reconstruction can be summarized as the following problem
\begin{equation}\label{E:IP:ML}
\begin{cases}
\min_{\bm{\Theta}}\ \mathbb{E}_{(\bu,\bmf)\sim \mathcal{P}}\ \ell(\mathcal{F}_{\bm{\Theta}}(\bmf), \bu)+r(\mathcal{F}_{\bm{\Theta}}),\qquad\mbox{supervised;}\cr
\min_{\bm{\Theta}}\ \mathbb{E}_{\bmf\sim \mathcal{P}}\ \ell(\bm{A}\mathcal{F}_{\bm{\Theta}}(\bmf), \bmf)+r(\mathcal{F}_{\bm{\Theta}}),\qquad\mbox{unsupervised.}
\end{cases}
\end{equation}
Here, $\mathcal{F}_{\bm{\Theta}}$ is an image reconstruction operator parameterized by $\bm{\Theta}$ that takes $\bmf$ as input and the reconstructed image as output, $\ell(\cdot,\cdot)$ is a certain loss function measuring the distance between the two input arguments, $r(\cdot)$ is a certain regularization on image reconstruction operator and $\mathcal{P}$ is the distribution of the data. We shall call \eqref{E:IP:ML} the learning-based approach (or deep learning approach when $\mathcal{F}_{\bm{\Theta}}$ is a DNN). Note that a possible variant for \eqref{E:IP:ML} is to consider the dependence of $\bm{\Theta}$ on $f$, i.e. $\mathcal{F}_{\bm{\Theta}(f;\bm{\Gamma})}$. Such formulation makes the model $\mathcal{F}_{\bm{\Theta}(f;\bm{\Gamma})}$ adaptive to each data $\bmf$ rather than having a fixed model $\mathcal{F}_{\bm{\Theta}}$ for all data in $\mathcal{P}$. This is closely related to meta-learning \cite{hospedales2021meta}.

The main difference between handcrafted modeling and learning-based approach (especially deep learning approach) is twofold: 
\begin{enumerate}
    \item the image reconstruction operator $\mathcal{F}_{\bm{\Theta}}$ is fully or mostly human-designed for handcraft modeling, while $\mathcal{F}_{\bm{\Theta}}$ is a complex nonlinear composite function with less human-designed structures but millions of trainable parameters ${\bm{\Theta}}$ for deep learning approach;
    \item the objective function in the optimization problem contains an expectation over a data distribution $\mathcal{P}$ for a learning-based approach, while the expectation is absent for typical handcraft models.
\end{enumerate}
The first difference leads to the well-known argument that DNNs and the training dynamics of \eqref{E:IP:ML} are generally harder to interpret than handcrafted modeling. The second difference is why learning-based models are more data-adaptive than handcrafted models, while it also raises the new issue of generalization since the expectation can only be approximated using samples. 

Both handcrafted models and deep learning models have their advantages and drawbacks depending on the applications. There has been an increasing effort in the community to combine handcrafted modeling and deep modeling to enjoy benefits from both approaches. One of the popular ways of such a combination is the so-called unrolled dynamics approach. It started with the seminal work of \cite{gregor2010learning} where the authors showed that one could unroll the iterative soft-thresholding algorithm (ISTA) to create a feed-forward network. Then, one can train ISTA end-to-end to determine the parameters in ISTA so that they are best suitable for the training data. The trained ISTA can achieve much smaller errors than the original ISTA or even FISTA \cite{beck2009fast}. This work initiated an approach to unroll a discrete dynamic system (an optimization algorithm or a differential equation) to form a network for end-to-end training. Recently, more and more examples showed that the unrolling dynamics approach well balances model interpretability and efficacy. This includes unrolling discrete forms of nonlinear diffusion equations for image restoration \cite{Liu2010Learning,chen2015learning} and unrolling optimization algorithms for image reconstruction and inverse problems \cite{sun2016deep,zhang2017learning,adler2018learned,solomon2019deep,chen2018theoretical,pmlr-v95-li18f,yang2018admm,zhang2019JRSNet,dong2019denoising,li2020efficient,zhang2020metainv}. The unrolling dynamics approach can often result in deep models that have better interpretability inherited from the original dynamics. Furthermore, these deep models normally have much fewer trainable parameters than black-box deep neural networks (e.g., using a plain CNN, U-Net, ResNet, etc.), which are more suitable for learning on relatively small data sets. We refer the interested readers to \cite{monga2021algorithm} for a comprehensive review on the unrolled dynamics approach.

The unrolled dynamics models often start with an iterative algorithm which can be an optimization algorithm or a discretization of a certain evolution PDE. For example, we can use the following iterative algorithm as the backbone dynamics
\begin{equation}\label{eq:abstract-BCD}
\begin{cases}
\bm{u}^{k+1}&=\arg\min_{\bm{u}}E_{\lambda}(\bm{u},\bm{z}^{k};\bm{\beta}^{k}),\cr
\bm{z}^{k+1}&=\arg\min_{\bm{z}}E_{\lambda}(\bm{u}^{k+1},\bm{z};\bm{\beta}^{k}),\cr
\bm{\beta}^{k+1}&= \bm{\beta}^{k}+\gamma(\bm{H}\bm{u}^{k+1}-\bm{z}^{k+1}),\qquad k=0,1,\ldots,K-1.
\end{cases}
\end{equation}
Here, the function $E_{\lambda,\gamma}(\bm{u},\bm{z};\bm{\beta})$ is defined as $$E_{\lambda,\gamma}(\bm{u},\bm{z};\bm{\beta})=\mathcal{D}(\bm{u},\bmf)+\bm{\gamma}\mathcal{C}_{\bm{H}}(\bm{u},\bm{z};\bm{\beta})+\lambda \mathcal{R}(\bm{z}),$$
where $\mathcal{D}(\bm{u},\bmf)$ is a data fidelity term (e.g., $\mathcal{D}(\bm{u},\bmf)=\frac12\|\bm{A}\bu-\bmf\|_2^2$), $\mathcal{C}_{\bm{H}}(\bm{u},\bm{z};\bm{\beta})$ is a constraint term that links the primal variable $\bm{u}$ with a certain sparsifying transformation $\bm{H}$ (e.g., $\mathcal{C}_{\bm{H}}(\bm{u},\bm{z};\bm{\beta})=\frac12\|\bm{H}\bu-\bm{z}+\bm{\beta}\|$), the auxiliary variable $\bm{z}$ and the dual variable $\bm{\beta}$, and $\mathcal{R}(\bm{z})$ is the regularization term (e.g., $\mathcal{R}(\bm{z})=\|\bm{z}\|_1$). The iterative algorithm \eqref{eq:abstract-BCD} defines a discrete dynamics. If it is unrolled, it forms a feed-forward network that takes $\bu^0$ as input and $\bu^{K}$ as output. It is common practice to choose $\bu^0$ as a coarse approximation of the image to be reconstructed computed from $\bmf$. Thus, this feed-forward network forms the nonlinear operator $\mathcal{F}_{\bm{\Theta}}$ in \eqref{E:IP:ML}. The parameter $\bm{\Theta}$ may include the hyperparameters $\lambda$ and $\gamma$ in \eqref{eq:abstract-BCD}, and any trainable parameters introduced in the $\bm{u}$- and $\bm{z}$-subproblem.


Most existing unrolled dynamics models have the tendency to define as many learnable components in the dynamics as possible, which makes the ultimate model further deviate from the dynamics it started with. Having a lot of learnable components increases the express power of the model to enable it to approximate more complex mappings. At the same time, however, its training can be more challenging and requires more data. In addition, an unrolled dynamic model with too many learnable components may also be susceptible to distribution shift during inference which is inevitable in practice. Therefore, we suggest an opposite direction in \cite{zhang2020metainv} that keeps the learnable components to the minimum and only converts the components to learnable units when they are most critical to the reconstruction quality and too intricate to be handcrafted.

Now, we briefly recall the model and main empirical findings of \cite{zhang2020metainv}. Consider the following optimization model for image reconstruction
\begin{equation}\label{E:HMZhang}
\min_{\bm{u},\bm{z}} \frac{1}{2}\|\bm{A}\bm{u}-\bm{f}\|^{2}+\sum_{\ell=1}^L\left(\lambda_i\|\bm{z}_\ell\|_{1}+\frac{\gamma_\ell}{2}\|\bm{W}_{\ell}\bm{u}-\bm{z}_{\ell}\|^2\right),
\end{equation}
where $\bm{W}_\ell$ is a certain sparsifying transform such as wavelet frame transform. The optimization problem \eqref{E:HMZhang} can be solved by the following alternative optimization strategy
\begin{equation*}
\begin{cases}
 \bm{u}^{k+1}&=\arg\min_{\bm{u}}\|\bm{A}\bm{u}-\bm{f}\|^{2}+\sum_{\ell=1}^{L} \gamma_{\ell}\|\bm{W}_{\ell}\bm{u}-\bm{z}^{k}_{\ell}\|^2, \cr
 \bm{z}^{k+1}_\ell&=\arg\min_{\bm{z}} \lambda_\ell \|\bm{z}_\ell\|_{1}+\frac{\gamma_\ell}{2} \|\bm{W}_{\ell}\bm{u}^{k+1}-\bm{z}_{\ell}\|^2,\quad\ell=1,\ldots,L,
 \end{cases}
\end{equation*}
with proper initialization $\bm{u}_{0}$ and $\bm{z}_{0}$. Solution to each of the two subproblem takes the form
\begin{equation}\label{E:HMZhang:UD}
\begin{cases}
 \bm{u}^{k+1}&=\left(\bm{A}^{\top}\bm{A}+\sum_{\ell=1}^{L}{\gamma}_{\ell}\bm{W}_{\ell}^{\top}\bm{W}_{\ell}\right)^{-1}\left[ \bm{A}^{\top}\bm{f}+\sum_{\ell=1}^{L}{\gamma}_{\ell}\bm{W}_{\ell}^{\top}\bm{z}_{\ell}^{k}\right], \cr
 \bm{z}^{k+1}_\ell&=\mathcal{T}_{\lambda_\ell/\gamma_\ell}(\bm{W}_\ell\bm{u}^{k+1}),\quad\ell=1,\ldots,L,
\end{cases}
\end{equation}
where $\mathcal{T}_\beta$ is the soft-thresholding operator with threshold level $\beta$. Note that the $\bm{u}$-subproblem of \eqref{E:HMZhang:UD} can be solved by the conjugate gradient (CG) method. The design of the deep unrolled dynamics model of \cite{zhang2020metainv} is to use a CNN as a hypernetwork that takes the current best approximation of $\bm{u}$ as input and initialization for the CG method as output at each step $k$ while keeping all other components manually selected. Most existing unrolled dynamics methods tend to convert the sparsifying transform, the soft-thresholding operator, or even the inversion of the linear system of \eqref{E:HMZhang:UD} to learnable units. However, \cite{zhang2020metainv} empirically demonstrated that they are all unnecessary, and we only need to use a CNN to infer a good initialization for the CG method that solves the $\bm{u}$-subproblem. Comprehensive experiments of \cite{zhang2020metainv} show that such design of the unrolled dynamics model is particularly beneficial in terms of robustness to distribution shift that may be introduced by using a different image data set, noise level/type, imaging conjuration, etc. 

There are a few remaining questions on the unrolled dynamics approach that are worth further exploration. Although empirical studies show that this approach leads to a model $\mathcal{F}_{\bm{\Theta}}$ more interpretable and generalizes better than some black-box deep learning models, there is still in lack of theoretical support to these empirical studies. Some pioneering works \cite{xu2019can,chen2020understanding} attempted to tackle this problem by analyzing the complexity and PAC-learnability of these networks. Nonetheless, the unrolled dynamics approach severely lacks theoretical guidance on model design and training. One advantage of the unrolled dynamics approach is the incorporation of our existing knowledge on the imaging modality and reconstruction algorithms. However, this may as well limit the expressive power of the model leading to only incremental improvements. Furthermore, the modeling of the underlying imaging modality with an operator $\bm{A}$ is only an approximation which can be a very rough one in practice. Thus, there are some works (e.g., \cite{zhu2018image}) that suggest using black-box deep learning models instead, hoping the model itself can automatically correct such inaccuracy. Of course, all of these are speculations and require further empirical and theoretical studies. 

While most of the models we talked about are within the scope of supervised learning, unsupervised (and self-supervised) learning is another prevailing approach for image reconstruction \cite{bora2017compressed,shah2018solving,ulyanov2018deep,van2018compressed,jagatap2019algorithmic,batson2019noise2self,krull2019noise2void,moran2020noisier2noise,quan2020self2self,pang2020self,pang2021recorrupted}. In comparison with the supervised learning-based approach, unsupervised methods are more reliant on the regularization term (i.e., $r(\cdot)$ in \eqref{E:IP:ML}) due to the lack of labels. Nonetheless, regularization is the key to the success of both supervised and unsupervised learning-based approaches since it can effectively reduce model complexity and improve generalization. However, unlike mathematical approaches studied extensively for decades, the regularization of the learning-based approach is still a mystery in general. Many effective regularizations are implicitly induced by the stochasticity of the data, model, and training algorithm \cite{gunasekar2018characterizing,ji2019implicit,vaskevicius2019implicit,chizat2020implicit,ali2020implicit,razin2020implicit}.  Therefore, a collective understanding of these regularization effects and guidance on exploiting them in practice is another important line of research.

\section{Improving Sensing with Reinforcement Learning}\label{SS:Learning2Scan}

In compressed sensing, the sensing strategy is random and not adaptive to the imaging subject \cite{candes2006robust,donoho2006compressed}, i.e., the measurements are randomly selected independently from different imaging subjects. In theory, such a strategy is proven for exact recovery using a convex model for only a handful of imaging modalities such as MRI. However, the theory of compressed sensing fails to cover many other imaging modalities, among which CT imaging is an important example due to the coherence structure of Radon transform. In practice, uniform sampling is often adopted due to its simplicity. However, for each subject, a uniform or random sensing strategy may not be ideal. It is more desirable to design a personalized sensing strategy for each subject to achieve better reconstruction results.  

In this section, we discuss how we may use reinforcement learning (RL) \cite{sutton2018reinforcement}, a powerful tool for sequential decision making, to improve sensing in computational imaging. In particular, we shall review the work of \cite{shen2020learning} for CT image reconstruction, where the we used RL to train a CT scanning policy that is adaptive to each imaging subject. Note that RL has also been used in other imaging modalities to improve sensing. For example, in Scanning Transmission Electron Microscopy (STEM), recent work by \cite{ede2021adaptive} proposes to use RL to guide the movement of the detector and uses a generator to generate reconstructed images. In \cite{pineda2020active}, RL is used to learn acquisition trajectories in MRI k-space for a fixed image reconstruction model. RL is also applied to improve autonomous exploration \cite{ly2019autonomous}.

In \cite{shen2020learning}, the scanning procedure is formulated as a Markov Decision Process (MDP), where the state includes currently collected measurements, the action determines the next measurement angle and the dose allocation, and the reward depends on the reconstruction quality measured by the peak signal-to-noise ratio (PSNR). Then, the scanning policy is trained using the proximal policy optimization (PPO) algorithm \cite{schulman2017proximal}. The key to the application of RL in sensing is the design of an appropriate MDP. Here, we review the MDP designed by \cite{shen2020learning} as follows.
\subsubsection*{MDP Formulation for Adaptive CT Scanning}
The CT scanning process can be viewed as a sequential decision process, where at each time step, we need to decide on the measurement angle and the corresponding X-ray dose. 
Given an Image $I$ and the number of all possible angles $N$ (e.g., $N = 360$ if we can choose all the integer angles from $0$\textdegree to $359$\textdegree), we now elaborate how the CT scanning process on $I$ can be formulated as an MDP:
\begin{itemize}
    \item[1)] The \textbf{state} is a sequence
    $\vec{s_t}=(s_1,s_2,...,s_t)$, where $s_t=(p_t, d_t^{\text{ac}},d_t^{\text{rest}})$. $p_t$ is the collected measurement at time step $t$. $d_t^{\text{ac}}$ records the used dose distribution up to time step $t$. The scalar $d_t^{\text{rest}}$ represents the amount of the remaining dose. 
    
    \item[2)] The \textbf{action} is $a_t=(a_{t}^\text{angle},a_{t}^{\text{dose}})$. $a_{t}^\text{angle}$ is a one-hot vector recording the angle we choose at time step $t$. $a_{t}^{\text{dose}} \in [0,1]$ is the fraction of dose that we apply at the corresponding angle. We terminate the MDP when the total used dose exceeds the total allowed dose. 
    
    \item[3)] The \textbf{reward} is computed as  $r(s_t, a_t) = \text{PSNR}(I_t, I)-\text{PSNR}(I_{t-1}, I)$, where $I$ is the groundtruth image, $I_t$ is the reconstructed image at time step $t$, and $\text{PSNR}(\hat{I}, I)$ represents the PSNR value of the reconstructed image $\hat{I}$. The reconstructed image $I_t$ is obtained by SART \cite{SART}.
    
    \item[4)] The \textbf{transition model} represents the scanning process of CT. At time step $t$, given the state $\vec{s}_t$ and action $a_t$, the next state $\vec{s}_{t+1}$ is simply the concatenation of $\vec{s}_t$ and $s_{t+1} = (p_{t+1}, d_{t+1}^{\text{ac}}, d_{t+1}^{\text{rest}})$. Details on how each of the three elements in $s_{t+1}$ is computed can be found in \cite{shen2020learning}. 
\end{itemize}
The policy network is chosen as a Recurrent Neural Network (RNN) to combine all the information from the past measurements. The RNN is also carefully designed to handle continuous action and discrete action simultaneously. We refer the interested readers to \cite{shen2020learning} for details. Numerical experiments on a large CT image dataset show that RL's scanning policy significantly outperforms the standard scanning strategies in terms of reconstruction quality. In contrast to \cite{ede2021adaptive}, the RL's scanning policy can be directly combined with different image reconstruction algorithms without retraining. 

Note that sensing for many imaging modalities (e.g., CT and MRI) can be formulated as combinatorial optimizations. In recently years, there is an emerging line of research that applies RL to solve combinatorial optimization problems (COPs) with exciting results, such as Travelling Salesman Problem \cite{bello2016neural}, Vehicle Routing Problem \cite{kool2018attention}, Influence Maximization \cite{mittal2019learning}, Autonomous Exploration \cite{auto_explore}, Integer Programming \cite{tang2020reinforcement}, etc. We can view RL as a trainable optimization algorithm that can continue improving itself by attempting to solve sampled COPs. However, the success of RL in solving a COP heavily relies on the design of the MDPs, the exploitation of the special structures of the COP in hand, and the choice of training algorithm and tricks. Nonetheless, RL is a promising approach for COPs and, in particular, can make sensing of computational imaging more data-adaptive and even task-driven (see the next section).

\section{Task-Driven Computational Imaging}\label{SS:End2End}

In this section, we discuss a possibility to unify the three steps, i.e., sensing, image reconstruction, and image analysis, of computational imaging. Although this is not a new idea, we can make it more practical and computationally tractable with the latest advancements in machine learning and its integration with mathematical approaches.

The motivation for making such unification in computational imaging is twofold. Firstly, it induces a task-specific quality metric for sensing and image reconstruction. Here, a task-specific quality metric is a metric to evaluate the quality of a reconstructed image in fulfilling a specific image analysis task. A proof-of-concept study regarding this was given by \cite{wu2018end}, where we showed that an image abnormality detection task may induce a significantly different image quality metric from pixel-level metric. Works in computer vision \cite{liu2018image,liu2020connecting} and radiology \cite{kalra2018radiomics} also suggest potentials of task-driven imaging, and a theoretical framework was introduced by \cite{adler2021task}. Therefore, the unification may lead to more effective and practically relevant ways to evaluating imaging qualities and more economic sensing strategies for a given task or a set of tasks.  Secondly, image analysis can benefit from such unification as well (see  \cite{wu2018end} for some empirical evidence). This is because the measurements acquired from sensing contain more information than the underlying reconstructed image due to inevitable information loss induced by any image reconstruction algorithm. 

On the technical side, the unification is conceptually straightforward while computationally challenging. We illustrate how the unification can be realized under the supervised learning regime. Let $\mathcal{M}_{\bm{\Theta}_1}: \bu \mapsto \bmf$ be a parameterized sensing operator that maps an image $\bu$ to its measurements $\bmf$; $\mathcal{F}_{\bm{\Theta}_2}: \bmf\mapsto\bu$ be a parameterized image reconstruction operator (e.g. the one defined in \eqref{E:IP:ML}); $\mathcal{G}_{\bm{\Theta}_3}: \bu\mapsto \bz$ be a parameterized image analysis operator that takes an image $\bu$ to its task-dependent value $\bz$ (e.g. one-hot vectors for image classification). Then, the unification can be achieved by solving the following optimization problem
\begin{equation}\label{E:Unification}
\min_{\bm{\Theta}_1, \bm{\Theta}_2, \bm{\Theta}_3}\ \mathbb{E}_{(\bu,\bz)\sim \mathcal{P}}\ \ell\left(\mathcal{G}_{\bm{\Theta}_3}\circ\mathcal{F}_{\bm{\Theta}_2}\circ\mathcal{M}_{\bm{\Theta}_1}(\bu)\right), \bz).
\end{equation}
Note that, unlike \eqref{E:IP:ML}, we have dropped the regularization terms on the three nonlinear operators for simplicity. 

The optimization problem \eqref{E:Unification} can be a very tough problem to work with. First of all, it is not clear whether there exists good parameterized approximations to sensing, image reconstruction and image analysis, each of which is a complicated HD mapping. Second challenge comes from solving the optimization of \eqref{E:Unification} which includes the sampling of $\mathcal{P}$ and the design of optimization algorithms. Latest advances of machine learning, especially deep learning, have provided us with new tools to tackle \eqref{E:Unification}. Both empirical \cite{Krizhevsky2012ImageNet,Ronneberger2015U-net,He2016Deep,goodfellow2014generative,Lecun2015Deep,deep-learning-book-2016,silver2016mastering} and theoretical studies \cite{siegel2021optimal,weinan2019priori,chen2019note,weinan2018exponential,lu2020deep,montanelli2019deep,shen2021deep,shen2021neural,yarotsky2019phase} showed that DNNs are very effective in approximating nonlinear operators in HD spaces. The combinatorial nature of the sensing operator $\mathcal{M}_{\bm{\Theta}_1}$ can be effective handled by RL (as discussed in Section \ref{SS:Learning2Scan}) or Monte Carlo methods (e.g. Monte Carlo tree search \cite{browne2012survey}). Furthermore, the back-propagation (BP) \cite{rumelhart1986learning} based stochastic optimization algorithms and the development of deep learning platforms such as Tensorflow \cite{abadi2016tensorflow} and Pytorch \cite{paszke2019pytorch} have made it easier to build and efficiently optimize complex models such as \eqref{E:Unification} on multiple processors in parallel. 

As a result, special instances and different versions of \eqref{E:Unification} has already been considered in the literature. For example, \cite{shen2020learning,pineda2020active} optimized only $\mathcal{M}_{\bm{\Theta}_1}$ of \eqref{E:Unification} with $\mathcal{F}_{\bm{\Theta}_2}$ fixed and $\mathcal{G}_{\bm{\Theta}_3}$ being an identity operator (with $\bz=\bu$); \cite{jin2019self,ede2021adaptive,yin2021end} optimized \eqref{E:Unification} for both $\mathcal{M}_{\bm{\Theta}_1}$ and $\mathcal{F}_{\bm{\Theta}_2}$ while $\mathcal{G}_{\bm{\Theta}_3}$ is taken as an identity operator; \cite{liu2018image,liu2020connecting,wu2018end,huang2019brain} optimized $\mathcal{F}_{\bm{\Theta}_2}$ and $\mathcal{G}_{\bm{\Theta}_3}$ of 
\eqref{E:Unification} with $\mathcal{M}_{\bm{\Theta}_1}$ fixed. In the rising field of intelligent sensor design \cite{ballard2021machine}, all three operators are considered except that $\mathcal{M}_{\bm{\Theta}_1}$ is often relatively simple. In another emerging field known as the ``deep optics" \cite{wetzstein2020inference}, the operator $\mathcal{M}_{\bm{\Theta}_1}$ represents an optical module (a physical layer) while $\mathcal{G}_{\bm{\Theta}_3}\circ\mathcal{F}_{\bm{\Theta}_2}$ is the digital image processing module. The optical and image processing modules are first end-to-end trained for a specific imaging task. Then, the optical module can be fabricated from the trained operator $\mathcal{M}_{\bm{\Theta}_1}$ using 3D printing \cite{chang2018hybrid}. This provides a framework for end-to-end task-driven camera design.

Although the aforementioned works suggested potentials of the unification \eqref{E:Unification}, its full power is yet to be uncovered. Furthermore, we know very little on the theoretical aspects of the problem \eqref{E:Unification}, e.g., sample complexity, approximation, generalization, robustness to distribution shift, etc. This leaves a tremendous room for both empirical and theoretical studies.

\bibliographystyle{plain}
\bibliography{references}

\end{document}